\documentstyle[psfig]{elsart}

\newcommand{\epr}{e-print astro-ph/}

\begin{document} 
\begin{frontmatter}

\title{Lorentz Invariance Violation and the Spectrum and Source Power of
Ultrahigh Energy Cosmic Rays}
\author{F.W. Stecker}

\address{NASA Goddard Space Flight Center, Greenbelt, MD 20771}
\author{S.T. Scully\thanksref{nsf}}
\address{James Madison University, Harrisonburg, VA 22807}
\thanks[nsf]{NASA Summer Faculty Fellow}
\begin{abstract}               
  
Owing  to their  isotropy,  it is  generally  believed that  ultrahigh
energy cosmic  rays (UHECRs) are  extragalactic in origin. It  is then
expected that  interactions of these  cosmic rays with photons  of the
cosmic background  radiation (CBR) should produce  a drastic reduction
in their  flux above  and energy  of about $5  \times 10^{19}$  eV (50
EeV), the so-called ``GZK effect''.  At present, the existence of this
effect is uncertain owing  to conflicting observational data and small
number  statistics.  We  show  here  that a  small  amount of  Lorentz
invariance   violation   (LIV), which   could   turn  off   photomeson
interactions of UHECRs with the  CBR, could explain the UHECR spectrum
as measured by {\it AGASA} which shows an excess of UHECRs at energies
above 100  EeV. If new results  from the {\it Auger}  array agree with
the {\it  AGASA} spectrum, this may  be interpreted as  evidence for a
small  amount of  LIV.  If,  on the  other hand,  the new  results are
consistent with  the {\it HiRes}  results favoring a GZK  effect, this
would place  severe constraints  on LIV and,  by implication,  on some
Planck  scale  quantum gravity  models.   We  also  discuss the  power
requirements  needed to  explain the  UHECR  spectrum for  a range  of
assumptions, including source  evolution and LIV and show  that in all
cases our results disfavor a $\gamma$-ray burst origin for the UHECRs.

\end{abstract}

\end{frontmatter}

\section{Introduction}

Because  of their  extreme energy  and isotropic  distribution,  it is
believed  that UHECRs  are extragalactic  in origin.   Observations of
UHECRs  having energies  in excess  of 10$^{20}$eV  (100  EeV) present
difficult puzzles with regard to their origin, their nature, and their
propagation   \cite{st03a}.   After  the   discovery  of   the  cosmic
background  radiation  (CBR),  Greisen  \cite{gr66} and  Zatsepin  and
Kuzmin  \cite{za66} pointed  out that  photomeson  interactions should
deplete the  flux of  cosmic rays with  energies above $\sim$  50 EeV.
One  of us \cite{st68},  using data  on the  energy dependence  of the
photomeson  production   cross  section,  then   made  a  quantitative
calculation of this ``GZK effect'' deriving the mean photomeson energy
loss attenuation  length for protons  as a function of  proton energy.
These results indicated  that the attenuation length of  a proton with
an energy greater than 100 EeV is less than 100 Mpc which is much less
than  the visible  radius of  the universe   (see  Figure \ref{eloss}
\cite{st68}).   Thus,  what  is  sometimes  referred  to  as  the  GZK
``cutoff'' is  not a true cutoff,  but a suppression  of the ultrahigh
energy cosmic ray flux arising from a limitation of proton propagation
through the cosmic background radiation.

More  recently, ground based  detectors have  detected a  double digit
number of events with energies above 100 EeV, placing the existence of
the GZK effect somewhat in doubt \cite{na00}-\cite{kn04}. This is also
because  of a  {\it prima  facie} conflict  between  the observational
results  \cite{ta03}, \cite{ab03}  and is  probably due,  at  least in
part, to poor statistics \cite{de03}.  The need to increase the number
of observed events in order  to improve the determination of the UHECR
spectrum,  particularly  above  100  EeV,  has resulted  both  in  the
development  of   the  more  sensitive  {\it   Auger}  detector  array
\cite{sw03}  and  the  proposal  of even  more  sensitive  space-based
detector  missions  such  as  {\it  EUSO} \cite{sc01}  and  {\it  OWL}
\cite{st04}.

\section{Violating Lorentz Invariance}  

With  the idea of  spontaneous symmetry  breaking in  particle physics
came  the suggestion  that  Lorentz invariance  (LI)  might be  weakly
broken at high energies  \cite{sa72}.  Although no true quantum theory
of gravity exists, it was suggested  that LI might be violated in such
a  theory with  astrophysical consequences  \cite{ac98}.  It  was also
suggested  that  a possible  natural  abrogation  of  the GZK  effect,
resulting  in  the  existence  of  significant  fluxes  of  UHECRs  at
trans-GZK  energies, could  be caused  by  a small  amount of  Lorentz
invariance  violation  (LIV)  \cite{sa72},  \cite{ki72},  \cite{co99}.
Recent  work has placed  strong constraints  on such  LIV, constraints
which  may have important  implications for  some quantum  gravity and
large extra dimension  models \cite{sg01}-\cite{ac04}. However, only a
very small  amount of LIV  could substantially curtail  the photomeson
production  and $e^+e^-$  pair production  that result  from  the high
energy protons interacting with the CBR.

In this  paper we investigate  the observational implications  of this
possible effect of a very small amount of LIV, {\it viz.}, that cosmic
rays could indeed reach us after originating at distances greater than
100  Mpc without undergoing  large energy  losses from  photomeson and
pair  production  interactions.  Such  a small  violation  of  Lorentz
invariance  might be  produced  by Planck  scale effects  \cite{al00},
\cite{ap03}.

\section{Kinematical Considerations}

A  simple  formulation  for  breaking   LI  by  a  small  first  order
perturbation  in  the  free  particle  Lagrangian  which  leads  to  a
renormalizable  treatment  has  been  given  by  Coleman  and  Glashow
\cite{co99}.   Using  this formalism,  these  authors  point out  that
different particles  can have different  maximum attainable velocities
(MAVs) which  can be different  from $c$.  If  we denote the MAV  of a
particle  of type $i$  by $c_{i}$  and the  difference $c_{i}  - c_{j}
\equiv  \delta_{ij}$ then  Coleman  and Glashow  have  shown that  for
interactions  of protons  with CBR  photons of  energy  $\epsilon$ and
temperature $T_{CBR} = 2.73$ K, pion production ($p+\gamma \rightarrow
N + \pi$'s) is kinematically forbidden to take place and thus {\it CBR
photomeson production interactions are turned off} if

\begin{eqnarray}
\delta_{p\pi} > 5 \times 10^{-24}(\epsilon/T_{CBR})^2.
\label{eq1}
\end{eqnarray}

The condition  for significantly increasing the threshold for 
electron-positron pair production interactions 
($p+\gamma\rightarrow p+e^+ + e^-$) is given by \cite{ap03}

\begin{eqnarray}
\delta_{ep} > {{(m_p + m_e)m_p}\over{E_f^2}}.
\label{eq2}
\end{eqnarray}

which is $\cal{O}$$(10^{-22})$ 
for a fiducial energy $E_f = 100$ EeV.
The relation given by eq. (2) is similar to that derived for $\gamma\gamma$
pair production in Ref. \cite{sg01}. 
These values are much  smaller than the constraint $\delta_{e\gamma} <
\sim10^{-15}$  found in  Ref. \cite{sg01} and also smaller than the
conditional constraint of $\delta_{ep}$ $<$ $\cal{O}$$(10^{-20})$  derived in 
Ref. \cite{ga04}. Thus, given  even  a very
small  amount of  LI  violation, both  photomeson and  pair-production
interactions of UHECR with the CBR can be turned off.

\section{Calculations}
\label{calc.sec}

We  consider a  power-law  source  spectrum that  will  be capable  of
explaining the  UHECR data  after propagation effects  are considered.
We  first   describe  our  model  for  propagating   the  high  energy
protons. We then calculate the effects of violating Lorentz invariance
on the photomeson and pair production energy loss mechanisms resulting
from the high  energy protons interacting with the  CBR. We derive and
show the  predicted spectra  to be observed  at the Earth  that result
from including LIV  effects in our models. We end  by giving the power
density  requirements of  the source  spectrum needed  to  produce the
observed UHECR flux.

We first reconsider  the standard picture for the  propagation of high
energy  protons, including energy  losses resulting  from cosmological
redshifting, pair production  and pion production through interactions
with cosmic background radiation  (CBR) photons (see above).  We shall
assume for this calculation a flat $\Lambda$CDM universe with a Hubble
constant   of   H$_0$   =    70   km   s$^{-1}$   Mpc$^{-1}$,   taking
$\Omega_{\Lambda}$ = 0.7 and $\Omega_{m}$ = 0.3. \cite{la04}
\footnote{Taking a flat Einstein-de Sitter universe with $\Lambda$ = 0
model does not significantly affect the results.}

\begin{figure}
  \centerline{\psfig{figure=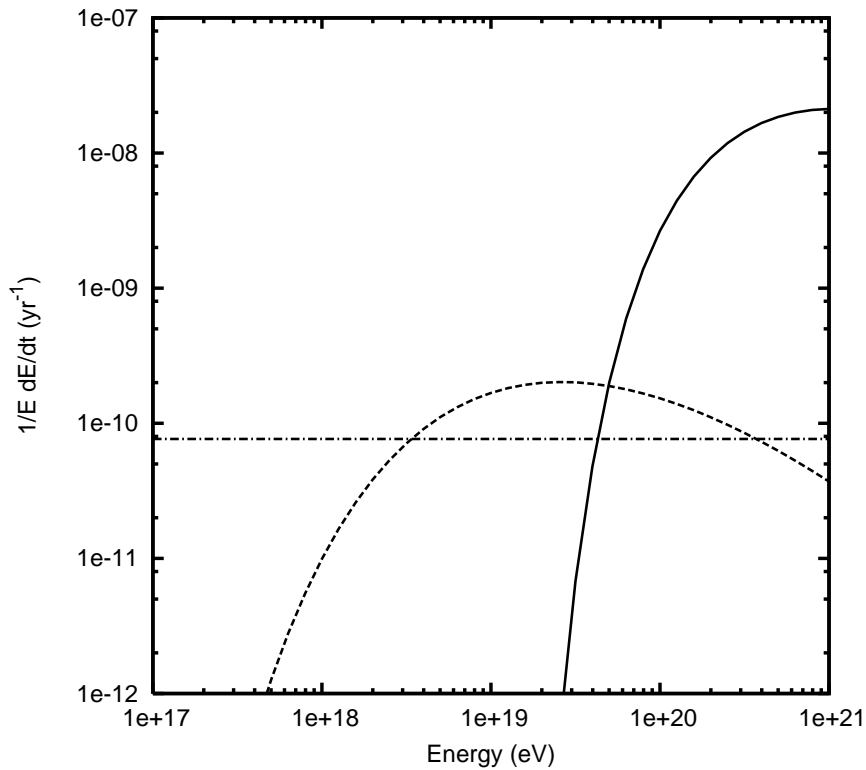,height=6.5in}}
\vspace{-1.5cm}
  \caption{Energy loss  rate curves for ultrahigh  energy protons from
photomeson  interactions   with  the  CBR   (solid),  pair  production
interactions with the CBR (dashed) and redshifting (dot-dashed).}
\label{eloss}
\end{figure} 

\vspace{1.5cm}

The energy  loss owing to  redshifting for a $\Lambda$CDM  universe is
then given by

\begin{eqnarray}
-(\partial \log  E/\partial t)_{redshift} =  H_{0}[\Omega_{m}(1+z)^3 +
  \Omega_{\Lambda}]^{1/2}.
  \label{eq3}
\end{eqnarray}

The combined  energy loss rate for  both pion and  pair production for
protons  in collisions with  photons of  the 2.7  K background  at the
present epoch ($z=0$) is defined as

\begin{eqnarray}
-(\partial  \log  E/\partial  t)_{\gamma  p}  \equiv  r_{\gamma  p}  =
r_{\pi}(E) + r_{e^+e^-}(E),
  \label{eq4}
\end{eqnarray}

where the energy loss rates $r(E)$ are defined at $z = 0$.  The photon
CBR number density increases as  $(1+z)^3$ and the CBR photon energies
increase  linearly with  $(1+z)$.  The corresponding  energy loss  for
protons at any redshift $z$ is thus given by
\begin{eqnarray}
r_{\gamma p}(E,z) = (1+z)^3 r[(1+z)E].
\label{eq5}
\end{eqnarray}
We take the photomeson  loss rate, $r_{\pi}(E)$, from Ref. \cite{st68}
and   the  pair-production  loss   rate,  $r_{e^+e^-}(E)$   from  Ref.
\cite{bl70}.  Figure \ref{eloss} shows the proton energy loss rates at
$z = 0$ as a function of energy.

Let $q(E_i,z)$ be the volume emissivity of particles produced by UHECR
sources at  redshift $z$ at an  initial energy $E_i$.   We assume that
$q(E_i,z)$ has a power-law energy dependence of the form

\begin{eqnarray}
q(E_i, z) = K(z)E_i^{-\Gamma}
  \label{eq6}
\end{eqnarray}

We further assume a redshift evolution of the volume emissivity of the
UHECR sources  to be of a form  similar to that of  the star formation
rate, {\it viz.},
\begin{eqnarray}
 K(z) = K(0)(1+z)^{(3+\zeta)}.
  \label{eq7}
\end{eqnarray}

\begin{figure}
  \centerline{\psfig{figure=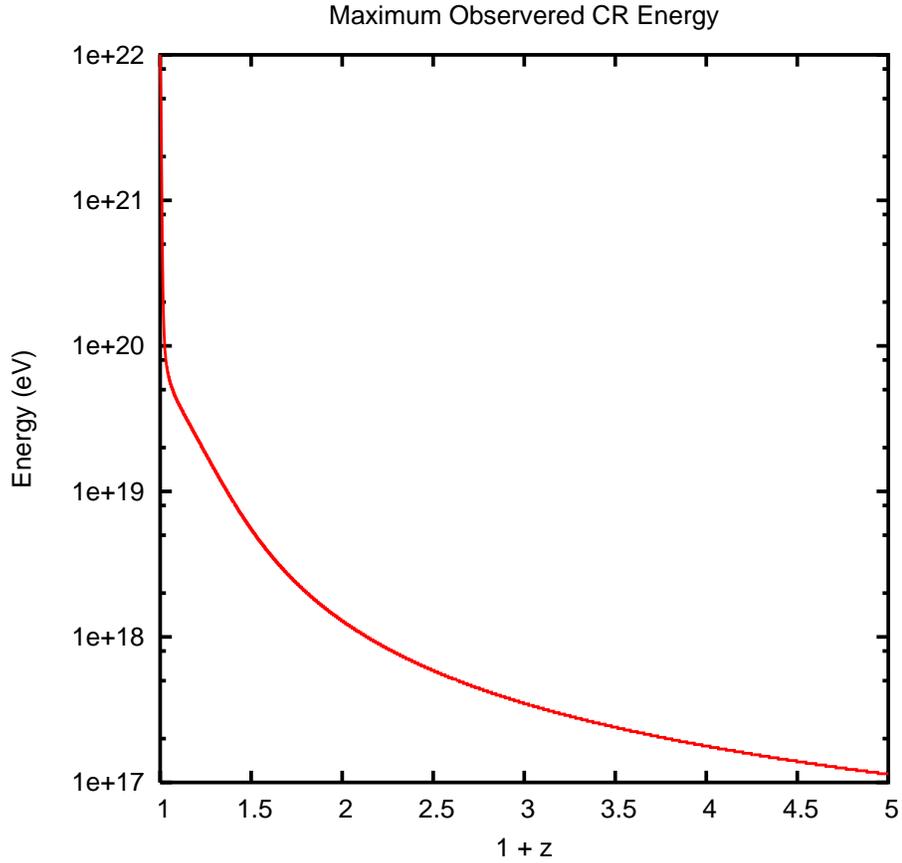,height=7.0in}}
\vspace{-1.5cm}
  \caption{The ``maximum'' energy a  proton should be observed with at
Earth ({\it  i.e.}, ``the  GZK cutoff energy'')  if it is  produced at
redshift $ z$.}
\label{egzk}
\end{figure} 
\vspace{1.5cm}

We first calculate  the initial energy $ E_i(z)$ at  which a proton is
created at a redshift $z$ whose observed energy today is $E$ following
the methods detailed in  Refs. \cite{be88} and \cite{sc02}. We neglect
the  effect of  possible small  intergalactic magnetic  fields  on the
paths  of these  ultrahigh energy  protons and  assume that  they will
propagate along straight lines from their source.  We then compute the
``GZK  energy''  for  UHECRs  versus  redshift  of  origin  using  the
$\Lambda$CDM model with the  cosmological parameters given above. This
is defined as  the energy at which the spectrum drops  to $1/e$ of its
original value  owing to photomeson  production.  We have  updated our
result from  Ref.  \cite{sc02}  to reflect the  $\Lambda$CDM cosmology
used in this  paper and the new result is  given in Figure \ref{egzk}.
It can be seen from Figure \ref{egzk} that the protons of energy above
10 EeV  can not  have originated  further than a  redshift of  $z \sim
0.4$.

\begin{figure}
  \centerline{\psfig{figure=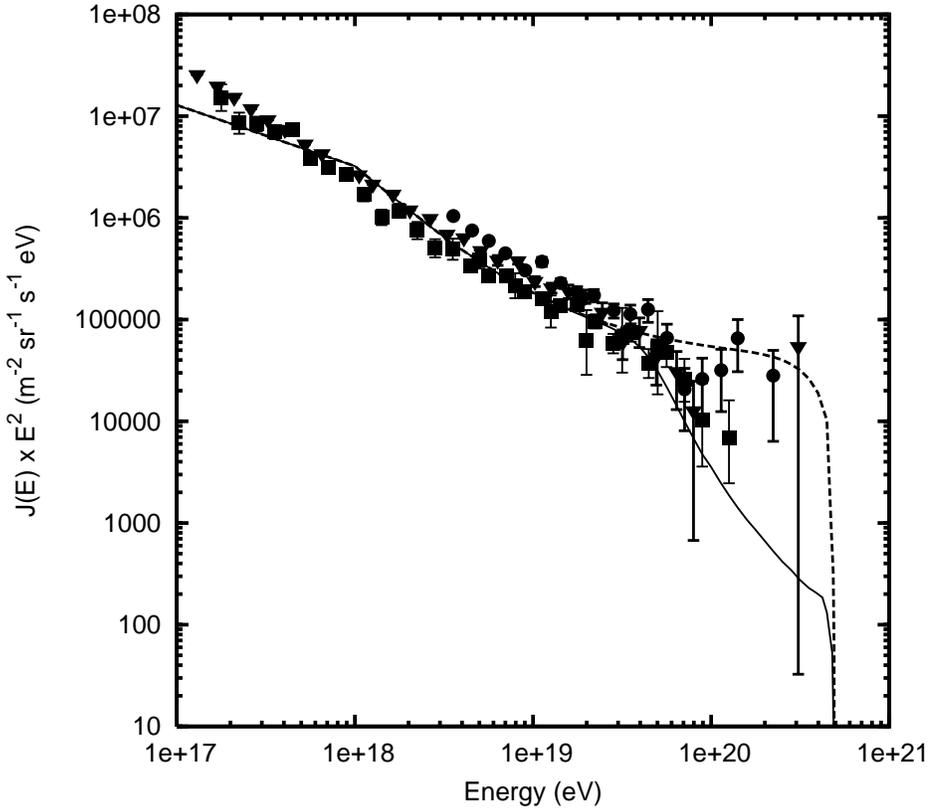,height=7.0in}}
\vspace{-1.5cm}
 \caption{Predicted  spectra for  an $E^{-2.6}$  source  spectrum with
source evolution (see text) shown with pair-production losses included
and photomeson  losses included (solid  curve) and turned  off (dashed
curve).   The  curves  are  shown  with ultrahigh  energy  cosmic  ray
spectral data from {\it Fly's Eye} (triangles), {\it AGASA} (circles),
and {\it HiRes} monocular data  (squares).  They are normalized to the
data at 3 EeV (see text).}
\label{f3}
\end{figure}
\vspace{1.5cm}

\begin{figure}
  \centerline{\psfig{figure=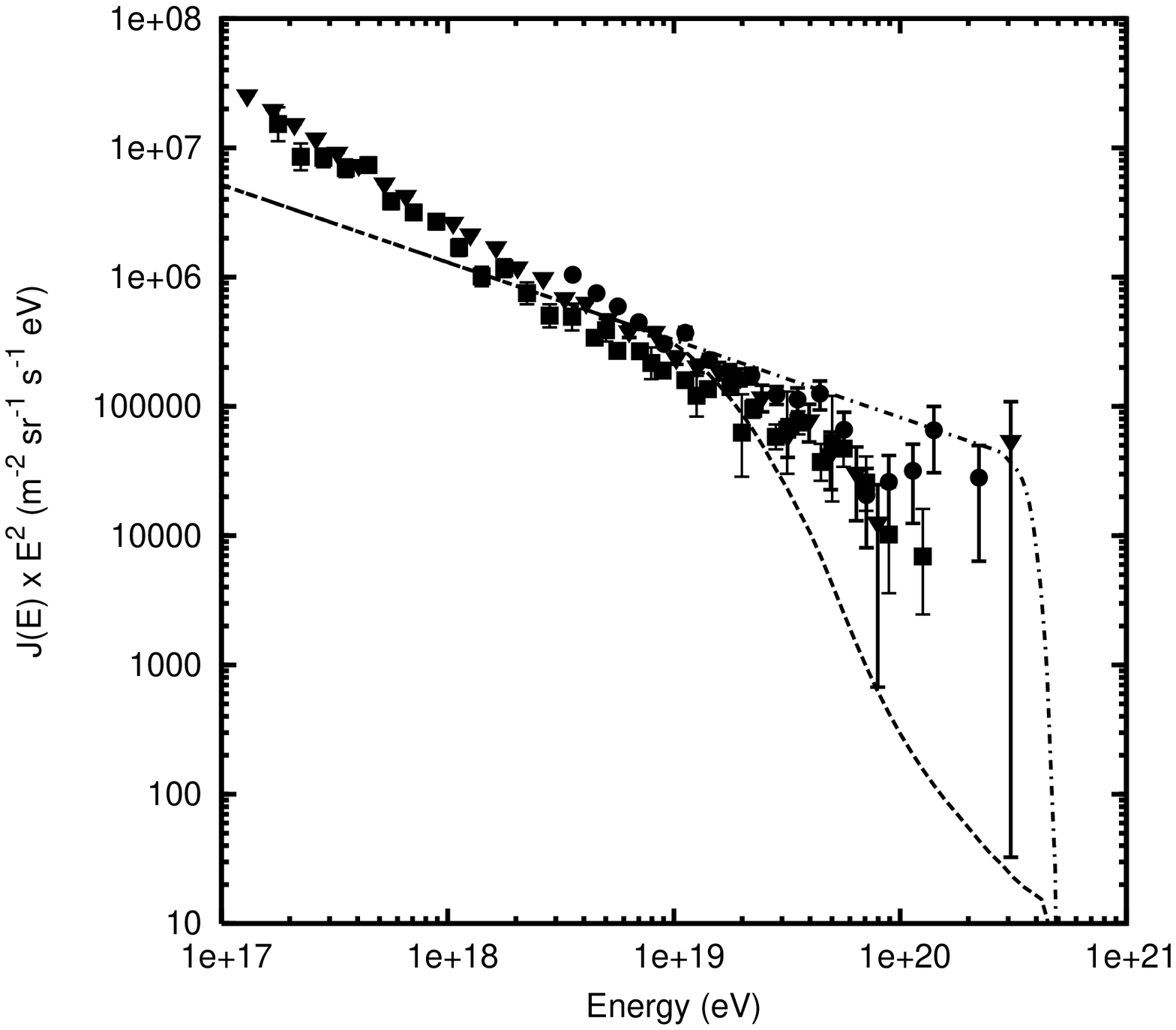,height=7.0in}}
\vspace{-1.5cm}
 \caption{Predicted  spectra for  an $E^{-2.6}$  source  spectrum with
source evolution  (see text) shown with  pair-production losses turned
off and with photomeson  production losses included (dashed curve) and
turned off  (dot-dashed curve).  The  curves are shown  with ultrahigh
energy cosmic ray spectral data from {\it Fly's Eye} (triangles), {\it
AGASA} (circles),  and {\it HiRes} monocular data  (squares). They are
normalized to the data at 3 EeV (see text.)}
\label{f4}
\end{figure}
\vspace{1.5cm}

The observed ultrahigh energy cosmic ray spectrum flattens above 3 EeV
\cite{bi94}, \cite{na00}.   This flattening is  usually interpreted as
being  caused by  the emergence  of a  harder  extragalactic component
which dominates  the spectrum at  energies above 3 EeV,  whereas below
this  energy the  observed spectrum  is  dominated by  a more  steeply
falling galactic component.

In our calculations  we assume a power law  extragalactic UHECR source
spectrum  with  a  value  for   $\Gamma$  of  2.6  which,  even  after
modification   by   energy    losses,   dominates   above   $\sim   3$
EeV.\footnote{We  have  tried using  a  flatter  source spectrum  with
$\Gamma = 2.35$,  but that this does not produce  a spectrum that fits
the observational data.}  For this  reason, and because the UHECR data
have smaller error bars at lower energies, we have chosen to normalize
our propagation-modified  extragalactic component  at 3 EeV.  Also, if
LIV turns off the GZK  effect, the maximum observable proton energy is
no longer limited by it.  Because the only significant energy loss for
protons with energy  above 500 EeV then comes from redshifting (see Figure
\ref{eloss}),  a power-law  source  spectrum will  produce a  power-law
UHECR  spectrum with  the  same spectral  index.   Therefore, for  the
purposes of  our calculations, we  assumed an arbitrary  maximum UHECR
energy of 500 EeV in the  source spectrum. However, in the case of LIV
with a  positive value of $\delta_{p\gamma} \sim$ $\cal{O}$(10$^{-23}$),
the maximum proton  energy can be limited by  the vacuum \v Cerenkov
effect \cite{co99},\cite{sg01}.

As in ref. \cite{sc02}, we have considered   two scenarios for the UHECR
emissivity  corresponding   to  different  values   for  the  redshift
evolution index,  $\zeta$, {\it viz.}, (1) a  constant comoving source
density distribution  ($\zeta = 0$),  {\it i.e.} no  source evolution,
out  to a  maximum  redshift, $z_{max}  =  6$, and  (2) an  emissivity
evolution  with  a  redshift   dependence  proportional  to  the  star
formation   rate,  taken   to   be  a   source  luminosity   evolution
corresponding  to $\zeta  =  3.6$ for  $0 <  z  < 2$  with no  further
evolution out to $z_{max} = 6$ \cite{bu04}.

\begin{figure}
  \centerline{\psfig{figure=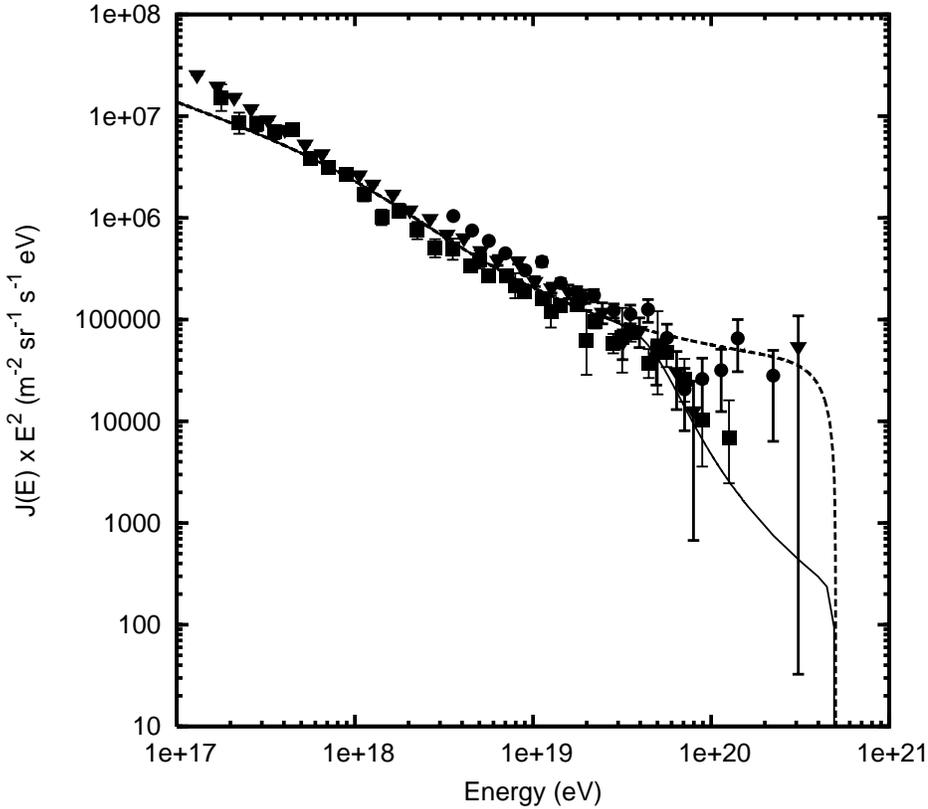,height=7.0in}}
\vspace{-1.5cm}  
 \caption{Predicted  spectra for an  $E^{-2.6}$ source  spectrum with
no evolution (see text) shown with pair-production losses included
and  photomeson losses included (solid  curve)  and turned  off
(dashed curve).  The curves are shown with ultrahigh energy cosmic ray
spectral data from {\it Fly's Eye} (triangles), {\it AGASA} (circles),
and {\it HiRes} monocular data  (squares).  They are normalized to the
data at 3 EeV (see text).}
\label{f5}
\end{figure}
\vspace{1.5cm}

We next incorporate the potential  effects of LIV on the proton energy
loss  rate into  our  model. In  the  first case (A) considered, this  is
implemented by  nullifying the energy loss  from photomeson production
as  a possible  consequence  of LIV.   In  addition to  this case,  we
present  the results obtained  for two  other possible  scenarios {\it
viz.}, (B) the case where  the pair production energy loss mechanism is
nullified but  not that from  photomeson production and, (C) the case
where  the energy  losses  from both  photomeson  production and  pair
production are nullified.   In all of these cases,  we have imposed an
arbitrary cutoff energy of 500  EeV on the postulated source spectrum.

\begin{figure}
  \centerline{\psfig{figure=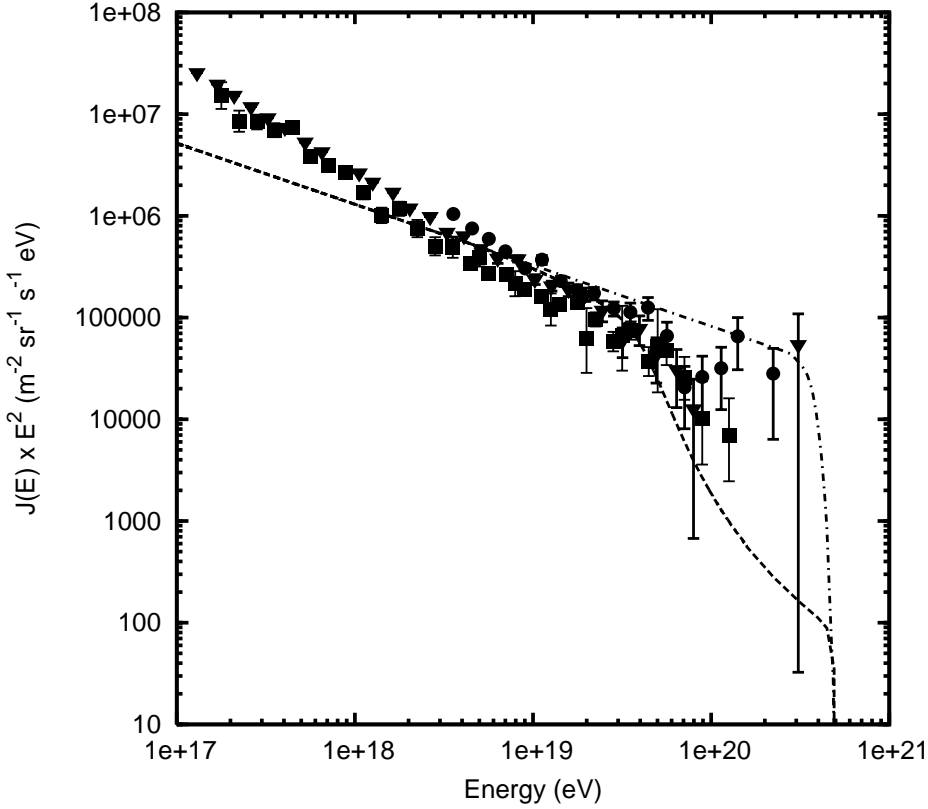,height=7.0in}}
\vspace{-1.5cm}  
 \caption{Predicted  spectra for an  $E^{-2.6}$ source  spectrum with
no evolution  (see text) shown with  pair-production losses turned
off and with photomeson  production losses included (dashed curve) and
turned  off (dot-dashed curve).  The curves  are shown  with ultrahigh
energy cosmic ray spectral data from {\it Fly's Eye} (triangles), {\it
AGASA} (circles),  and {\it HiRes} monocular data  (squares). They are
normalized to the data at 3 EeV (see text.)}
\label{f6}
\end{figure}
\vspace{1.5cm}

\section{Conclusions}

The  results of  our calculations, along with the spectra obtained
for the standard LI case,  are given  in Figures  \ref{f3} and \ref{f4}
for the source evolution model and Figures  \ref{f5} and \ref{f6} for
the no evolution ($\zeta = 0$) model. All of these figures also show
the  {\it HiRes}  \cite{ab03},  {\it Fly's  Eye}
\cite{fly94}  and {\it  AGASA} \cite{ta03}  data.\footnote{Other UHECR
data are given in Ref. \cite {kn04} and Ref. \cite{na00}.}
These spectra  are normalized  at an energy  of 3 EeV  above which
energy  the  extragalactic  cosmic  ray  component is  assumed  to  be
dominant  and below  which  we  assume that  the  galactic cosmic  ray
component dominates (see, {\it  e.g.}, Ref.  \cite{st03a}. The results
shown  are calculated  for ``on''  and ``off''  energy losses  for both
photomeson production and pair production  for protons. As can be seen
in Figures \ref{f3} and \ref{f5},  the LI case fits the {\it HiRes} spectrum, 
while the LIV case fits the {\it AGASA} spectrum.

It can also be seen from Figs. \ref{f3} - \ref{f6} that, in principle, a
very   small  amount   of   LI  violation   can   eliminate  the   GZK
``cutoff''. Also,  when pair-production is  turned off, the  $\sim$ 10
EeV ``bite'' in the predicted  spectrum is eliminated. Of course, when
both interactions are turned off, all of the features in the predicted
spectrum  disappear and  only  a power-law  remains.  Contrary to  the
discussion  of Ref. \cite{ap03},  as can  be seen  from the figures, 
the present  data cannot be  used to put constraints on LI 
violation in pair-production interactions.

The  UHECR source power  density requirements for  the source
evolution and  no evolution cases are different.  The absolute minimum
power  density required is  found by  integrating the  source spectrum
times particle energy from 3 EeV  to 500 EeV. Since we have normalized
this spectrum at 3 EeV  where photomeson production is unimportant, we
only have four cases to consider:

(1) With source evolution and no  pair production losses, we find that
the  minimum local ($z  = 0$)  power density  required is  $1.2 \times
10^{30}$ W Mpc$^{-3}$.

(2) With  source evolution  and including  pair production  losses, we
find  that the  local minimum  power density  required is  $1.5 \times
10^{31}$ W Mpc$^{-3}$.

(3) With no evolution and no  pair production losses, we find that the
local  minimum  power  density  required  is $7.7  \times  10^{30}$  W
Mpc$^{-3}$.

(4) With no  evolution and including  pair production losses,  we find
that the local minimum power  density required is $2.2 \times 10^{31}$
W Mpc$^{-3}$.

In all cases, we note that the local minimum power density required is
orders  of magnitude  larger than  that estimated  if the  sources are
$\gamma$-ray bursts and if we equate  their power output  in UHECRs with
their  power  output in  $\gamma$-rays.  {\it  viz.},  $\sim$$2 \times
10^{28}$  to  $\sim$$7 \times  10^{28}$  W  Mpc$^{-3}$  \cite{st00},
\cite{ms01}.  Of course,  the extragalactic component should naturally
extend  below 3 EeV.   If it  does so  with $\Gamma  = 2.6$,  this can
further increase the source energy requirements.

We consider
the source evolution case to  be a more probable scenario as the
frequency  of astrophysical production  sites, regardless  of specific
mechanism, is likely to be closely  tied to the star formation rate for
various reasons. 

If the GZK effect is confirmed, eq. (1) places a very strong constraint
on the LIV parameter $\delta_{p\pi}$. If the pair-production ``bite'' is
also clearly exhibited in data with better statistics, eq. (2) then
similarly constrains $\delta_{ep}$.  

If, on the other hand, LIV is the explanation for a missing  GZK effect, 
one  can also
look for  the absence  of a potential  ``pileup'' spectral  feature at
$\sim  70$ EeV  \cite{st89} and  for the  absence of  ultrahigh energy
photomeson  neutrinos.  Such  additional  future  observational  tests
should decide the issue.

\section*{Acknowledgment}
Work by FWS was supported in part by NASA Grant ATP03-0000-0057.
Work by STS was supported in part by a NASA Summer Faculty Fellowship.

\end{document}